\documentclass[%
 aip,
 amsmath,amssymb,
 reprint,%
]{revtex4-1}
\usepackage{color}
\usepackage[usenames,dvipsnames,svgnames,table]{xcolor}
\usepackage[colorlinks=true,linkcolor=blue,urlcolor=blue,citecolor=blue]{hyperref}
\usepackage{mathtools}
\usepackage{graphicx}
\usepackage{dcolumn}
\usepackage{array}
\usepackage{lipsum}
\usepackage{bm}
\usepackage{subfigure}
\usepackage{amssymb}
\usepackage{multirow}
\usepackage{tabularx}
\usepackage{amsmath}
\usepackage{braket}
\usepackage{csquotes}
\graphicspath{{plots/}}
 \usepackage{lipsum}
\usepackage{mathrsfs}
\usepackage{MnSymbol}
\usepackage{siunitx}
\newcommand{\beq}{\begin{equation}}
\newcommand{\eeq}{\end{equation}}
\newcommand{\bea}{\begin{eqnarray}}
\newcommand{\eea}{\end{eqnarray}}

\begin{document}
%\underline{Draft version:}
\title{Towards Model-free Temperature Diagnostics of Warm Dense Matter\\ from Multiple Scattering Angles}

\author{H.~M.~Bellenbaum}
\email{h.bellenbaum@hzdr.de}
\affiliation{Center for Advanced Systems Understanding (CASUS), D-02826 G\"orlitz, Germany}
\affiliation{Helmholtz-Zentrum Dresden-Rossendorf (HZDR), D-01328 Dresden, Germany}
\affiliation{Institut f\"ur Physik, Universit\"at Rostock, D-18057 Rostock, Germany}
\affiliation{Lawrence Livermore National Laboratory (LLNL), California 94550 Livermore, USA}

\author{B.~Bachmann}
\affiliation{Lawrence Livermore National Laboratory (LLNL), California 94550 Livermore, USA}

\author{D.~Kraus}
\affiliation{Institut f\"ur Physik, Universit\"at Rostock, D-18057 Rostock, Germany}
\affiliation{Helmholtz-Zentrum Dresden-Rossendorf (HZDR), D-01328 Dresden, Germany}

\author{Th.~Gawne}
\affiliation{Center for Advanced Systems Understanding (CASUS), D-02826 G\"orlitz, Germany}
\affiliation{Helmholtz-Zentrum Dresden-Rossendorf (HZDR), D-01328 Dresden, Germany}

\author{M.~P.~B\"ohme}
\affiliation{Center for Advanced Systems Understanding (CASUS), D-02826 G\"orlitz, Germany}
\affiliation{Helmholtz-Zentrum Dresden-Rossendorf (HZDR), D-01328 Dresden, Germany}
\affiliation{Technische  Universit\"at  Dresden,  D-01062  Dresden,  Germany}
\affiliation{Lawrence Livermore National Laboratory (LLNL), California 94550 Livermore, USA}

\author{T.~D\"oppner}
\affiliation{Lawrence Livermore National Laboratory (LLNL), California 94550 Livermore, USA}

\author{L.~B.~Fletcher}
\affiliation{SLAC National Accelerator Laboratory, Menlo Park California 94309, USA}

\author{M.~J.~MacDonald}
\affiliation{Lawrence Livermore National Laboratory (LLNL), California 94550 Livermore, USA}

\author{Zh.~A.~Moldabekov}
\affiliation{Center for Advanced Systems Understanding (CASUS), D-02826 G\"orlitz, Germany}
\affiliation{Helmholtz-Zentrum Dresden-Rossendorf (HZDR), D-01328 Dresden, Germany}

\author{Th.~R.~Preston}
\affiliation{European XFEL, D-22869 Schenefeld, Germany}

\author{J.~Vorberger}
\affiliation{Helmholtz-Zentrum Dresden-Rossendorf (HZDR), D-01328 Dresden, Germany}

\author{T.~Dornheim}
\email{t.dornheim@hzdr.de}
\affiliation{Center for Advanced Systems Understanding (CASUS), D-02826 G\"orlitz, Germany}
\affiliation{Helmholtz-Zentrum Dresden-Rossendorf (HZDR), D-01328 Dresden, Germany}

\begin{abstract}
% \textcolor{red}{\textbf{We need to process the backscattering data for the same shot numbers as Ben's forward scattering data to have a real apple to apple comparison. Hannah's summer internship project.}}
Warm dense matter (WDM) plays an important role in astrophysical objects and technological applications, but the rigorous diagnostics of corresponding experiments is notoriously difficult.
In this work, we present a model-free analysis of x-ray Thomson scattering (XRTS) measurements at multiple scattering angles.
Specifically, we analyze scattering data that have been collected for isochorically heated graphite at the Linac Coherent Light Source (LCLS).
Overall, we find good consistency in the extracted temperature between small and large scattering angles, whereas possible signatures of non-equilibrium may be hidden by the source function, and by the available dynamic spectral range.
The present proof-of-principle study directly points to improved experimental set-ups for equation-of-state measurements and for the model-free study of relaxation times.
% \todo{Hannah: Add synthetic analysis of forward scattering data from MCSS and a bit more detail here.}
\end{abstract}

\maketitle

%\section{Introduction\label{sec:introduction}}

The accurate understanding of matter under extreme densities, pressures, and temperatures constitutes a highly active frontier at the interface of plasma physics, laser science and quantum chemistry~\cite{drake2018high,wdm_book,fortov_review}.
In nature, these conditions abound in astrophysical objects such as giant planet interiors~\cite{Benuzzi_Mounaix_2014}, brown dwarfs~\cite{becker} and the outer layer of neutron stars~\cite{Haensel}. On Earth, such \emph{warm dense matter} (WDM) plays an important role in cutting-edge technological applications such as material synthesis and discovery~\cite{Kraus2016,Kraus2017,Lazicki2021}, and hot-electron chemistry~\cite{Brongersma2015}. A particularly prominent example is given by inertial confinement fusion (ICF)~\cite{Betti2016}, where the fuel capsule has to traverse the WDM regime on its compression path towards ignition~\cite{hu_ICF}.

Consequently, WDM is nowadays heavily studied in experiments using different techniques, see e.g.~the overview by Falk~\cite{falk_wdm}. On the one hand, this has led to a number of important achievements, such as the pioneering observation of plasmons in warm dense beryllium at OMEGA~\cite{Glenzer_PRL_2007} and the very recent in-depth experimental observation of partial K-shell ionization at the National Ignition Facility (NIF) also for Be~\cite{Tilo_Nature_2023}. On the other hand, the extreme conditions render the rigorous diagnostics of these extreme states a difficult challenge. Indeed, even basic parameters like temperature or density generally cannot be measured directly, and, instead, have to be inferred indirectly from other observations.

In this situation, X-ray Thomson scattering (XRTS) has emerged as a commonly used diagnostic method~\cite{siegfried_review} as it is, in principle, capable to give one access to the equation of state (EOS) of a given material~\cite{Gregori_PRE_2003,Falk_PRL_2014,MacDonald_POP_2023}.
More specifically, the measured scattering intensity is usually expressed as~\cite{Dornheim_T2_2022}
\begin{eqnarray}\label{eq:convolution}
    I(q,E) = S(q,E_0-E) \circledast R(E)\ ,
\end{eqnarray}
i.e., as a convolution of the dynamic structure factor $S(q,\hbar\omega)$ with the combined source and instrument function (SIF) $R(E)$ that takes into account both the shape of the probing x-ray source and detector effects~\cite{gawne_SIF}. Here $E$ and $E_0$ are the energy of the scattering photon and the incident beam energy, and the energy change is $\hbar \omega=(E-E_0)$. The corresponding momentum transfer is to a very good approximation only a function of the scattering angle $\theta$ at the conditions considered in the present work~\cite{Dornheim_T2_2022}.
This implies that XRTS does not give direct access to the physical information about the system of interest as the deconvolution that is required to solve Eq.~(\ref{eq:convolution}) for $S(q,\hbar\omega)$ is generally rendered highly unstable by the presence of noise in the experimental data. We note that even in cases where the deconvolution has been assumed possible~\cite{Sperling_PRL_2015}, its outcome and interpretation remains controversial~\cite{Mo_PRL_2018}.

In practice, the interpretation of a given XRTS data set is thus usually based on the construction of a forward model $S_\textnormal{model}(q,\hbar\omega)$ for the dynamic structure factor, which is subsequently convolved with the SIF and then compared with the experimental observation~\cite{Gregori_PRE_2003,siegfried_review,Tilo_Nature_2023,boehme2023evidence,kraus_xrts}.
Unfortunately, no reliable method that is capable of giving an exact description over the entire WDM parameter space exists, and one has to rely on de-facto uncontrolled approximations such as the ad-hoc decomposition into effectively \emph{free} and \emph{bound} electrons within the popular Chihara model~\cite{Chihara_1987,Gregori_PRE_2003,kraus_xrts}. As a consequence, the quality of the extracted EOS generally remains unclear.
First-principles simulations such as time-dependent density functional theory (TD-DFT)~\cite{dynamic2,Schoerner_PRE_2023,Moldabekov_PRR_2023} can, in principle, improve this situation, but this comes at the cost of a drastically increased computational demand that might render required parameter scans unfeasible.
In any case, it is clear that the forward-modelling based analysis of $I(q,E)$ constitutes a difficult inverse problem by itself, and the inferred conditions (e.g., temperature $T$ or mass density $\rho$) might not be unique~\cite{Kasim_POP_2019}.
Finally, the utility of different methods of generating WDM samples in the laboratory to probe the EOS may be limited by the presence of non-equilibrium~\cite{ernstorfer} and/or inhomogeneity effects~\cite{Chapman_POP_2014}.

Recently, it has been suggested to analyze the two-sided Laplace transform of the scattering intensity~\cite{Dornheim_T_2022,Dornheim_T2_2022},
\begin{eqnarray}\label{eq:Laplace}
    \mathcal{L}\left[I(q,E)\right] = \int_{-\infty}^\infty \textnormal{d}E\ I(q,E) e^{-E\tau}\ .
\end{eqnarray}
In combination with the well-known convolution theorem $\mathcal{L}\left[S(q,E)\circledast R(E)\right] = \mathcal{L}\left[S(q,E)\right] \mathcal{L}\left[ R(E)\right]$, Eq.~(\ref{eq:Laplace}) gives one direct access to the dynamic structure of the system \emph{in the imaginary-time domain},
\begin{eqnarray}\label{eq:ITCF}
    F(q,\tau) = \mathcal{L}\left[S(q,E)\right] = \frac{\mathcal{L}\left[S(q,E)\circledast R(E)\right] }{\mathcal{L}\left[ R(E)\right]}\ .
\end{eqnarray}
In particular, $F(q,\tau)$ corresponds to the usual intermediate scattering function $F(q,t)$, but evaluated for an imaginary-time argument $t=-i\hbar\tau$, with $\tau\in[0,\beta]$ and $\beta=1/k_\textnormal{B}T$ the inverse temperature in energy units. This imaginary-time correlation function (ITCF) naturally emerges in Feynman's powerful imaginary-time path integral framework for statistical mechanics~\cite{kleinert2009path} and, by definition, contains the same information as $S(q,E)$, only in an unfamiliar representation~\cite{Dornheim_insight_2022,Dornheim_PTR_2022,Dornheim_review}.
For example, the detailed balance relation~\cite{quantum_theory,DOPPNER2009182}
$S(q,-E) = S(q,E) e^{-E\beta}$ connects the up-shifted part of the scattering intensity where the scattered photon has gained energy from the system with the  down-shifted side that describes a corresponding energy loss. In thermal equilibrium, it gives one, in principle, access to the temperature of the probed system without the need for explicit models or approximations. In practice, its direct application to scattering data is usually prevented by the convolution with the SIF, see Eq.~(\ref{eq:convolution}).
The latter, however, is not a problem in the imaginary-time domain, where the detailed balance manifests as the symmetry relation~\cite{Dornheim_T_2022}
\begin{eqnarray}\label{eq:symmetry}
    F(q,\tau) = F(q,\beta-\tau)\ ,
\end{eqnarray}
which implies that $F(q,\tau)$ is symmetric around $\tau=\beta/2$, where it attains a minimum.
In other words, switching to the imaginary-time domain gives one model-free access to the temperature of arbitrarily complicated systems in thermodynamic equilibrium. Subsequently, this idea has been extended to infer other properties such as the electronic static structure factor~\cite{dornheim2023xray}, various frequency moments of $S(q,E)$~\cite{Dornheim_moments_2023}, and the Rayleigh weight $W_R(q)$ that describes electronic localization around the ions~\cite{dornheim2024modelfreerayleighweightxray}.

An additional strength of the Laplace method is its utility for the model-free detection of general non-equilibrium effects~\cite{Vorberger_PRX_2023}. Simply put, Eq.~(\ref{eq:symmetry}) only holds in thermal equilibrium and any deviation from this relation within the given confidence interval implies non-equilibrium effects if other sources of error such as the characterization of the SIF~\cite{MacDonald_POP_2022} can be excluded. 
In practice, this method is substantially more robust if one has access to multiple scattering angles.
For example, by collecting the scattering intensity both in a forward (FWD) and backward (BCK) direction, one effectively probes the system on large (collective) and small (single-particle) length scales~\cite{Dornheim_Nature_2022}.
In thermal equilibrium, this should result in the same set of free parameters as there is only a single density and temperature in the system, even though different angles are likely more sensitive to different parameters.
In contrast, Vorberger \emph{et al.}~\cite{Vorberger_PRX_2023} have shown that Eq.~(\ref{eq:symmetry}) is strongly violated even in the presence of a small fraction of non-thermal electrons based on synthetic spectra.

In the present work, we apply this idea for the first time to real scattering spectra that have been measured for isochorically heated graphite at the Linac Coherent Light Source (LCLS).
We note that such a self-scattering set-up constitutes a particularly interesting example for multiple reasons.
First, isochoric heating means that the density is a-priori known, which reduces the set of free parameters potentially simplifying the inference of an EOS~\cite{kraus_xrts}.
Second, x-ray pumping is both efficient in terms of energy deposition and can be expected not to induce strong temperature gradients due to partial heating.
Third, the measured scattering intensity constitutes a time-average over different conditions during the x-ray heating process, resulting in a non-equilibrium signal.
The degree of non-equilibrium thus crucially determines the value of self-scattering to learn something about the true EOS of WDM.

\textbf{Results.} 
Two different data sets were analysed here, with the experimental setup remaining the same (beam energy $E_b = 5.9\,\unit{eV}$, laser energy $E_L = 3\, \unit{mJ}$ and focus spot size $\leq 5 \, \unit{\mu m}$).
Two detectors were placed at $29$ and $160$ degrees to measure a forward and backward signal respectively.
A small CSPAD camera coupled to a HAPG crystal was used for the backward signal, while the forward spectrum was measured using a similar HAPG crystal and a PI camera \cite{kraus_xrts}.
The input spectrum was measured using a Silicon crystal and YAG detector upstream.
The first dataset analysed was averaged over 7 shots in the forward and 3 shots in the backward direction, since the remaining backscattering spectra from that run were unavailable.
The second was averaged over $\approx$ 40 shots taken during the same run.
No backward signal was recorded during this run.

\begin{figure}
    \centering
    \includegraphics[width=0.5\textwidth,keepaspectratio]{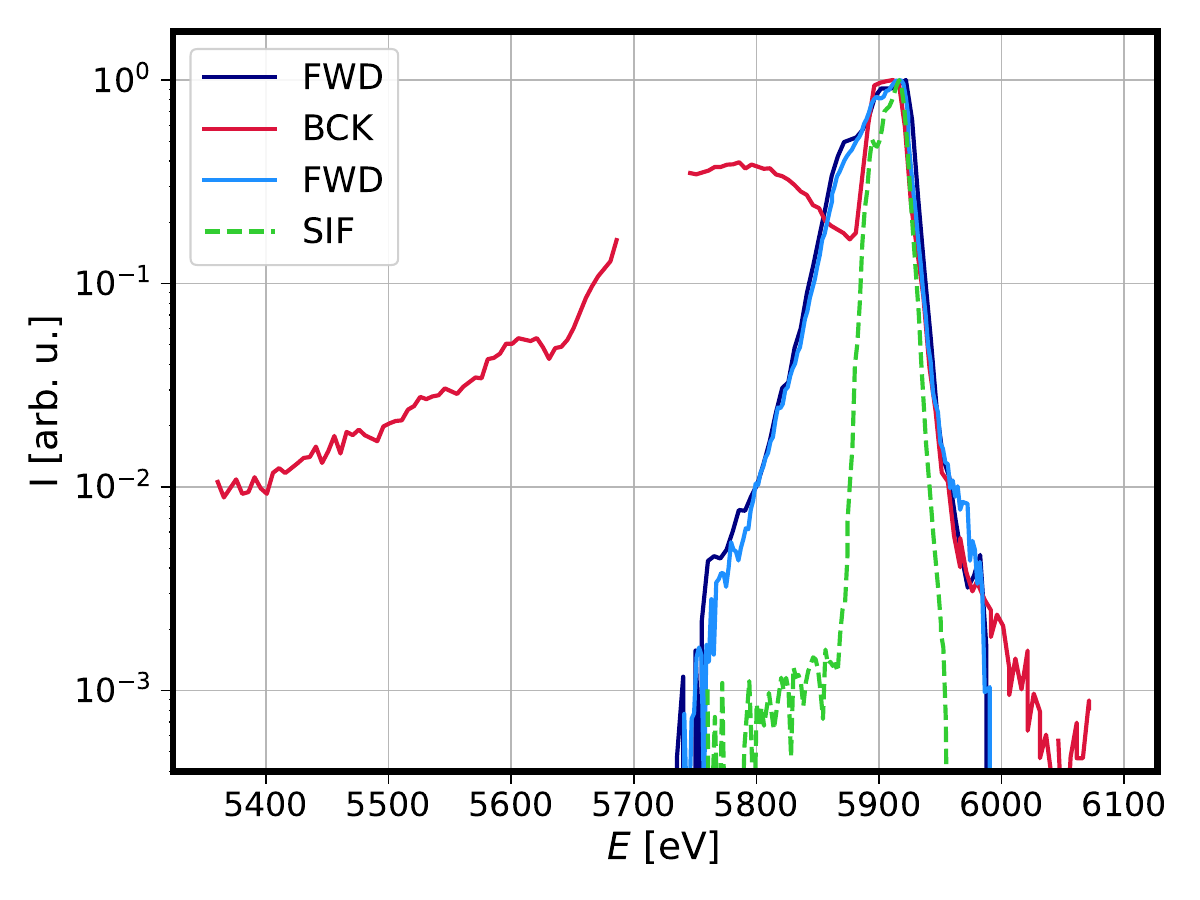}
    \caption{XRTS intensity as a function of photon energy. 
    Solid dark blue: forward scattering at $\theta = 29^\circ$; solid light blue: forward scattering taken from the second dataset for $\theta=29^\circ$; solid red: backward scattering at $\theta = 160^\circ$; dashed green: source spectrum.
    The source spectrum was arbitrarily scaled to match the measured spectra.
    The SIF plotted here is used in the forward analysis for the second dataset.
    Spectra were averaged over different runs taken at the same conditions, a corresponding source spectrum was used for each.
    }
    \label{fig:all-spectra}
\end{figure}

In Fig.~\ref{fig:all-spectra}, we show the measured scattering intensities that have been obtained by averaging over 3 independent shots, resulting in a dynamic range of over three orders of magnitude.
The backscattering signal is given by the red line; on the down-shifted side, it extends significantly beyond the K-edge, although there is a gap around $\hbar \omega=-200\,$eV due to a malfunctioning detector segment.
On the up-shifted side, we obtain a significant signal beyond $\omega=100\,$eV, indicating the presence of hot electrons.
In Ref.~\cite{kraus_xrts}, Kraus \emph{et al.}~have inferred a nominal temperature of $T=21.7\,$eV based on a Chihara model fit, although this estimate is subject to considerable uncertainties as the actual SIF (green line in Fig.~\ref{fig:all-spectra}) might have additional wings.
Subsequently, B\"ohme \emph{et al.}~\cite{boehme2023evidence} have shown that a temperature of $T=16.6\,$eV is more realistic if one takes into account the physically mandated, but previously neglected contributions of free-bound transitions; this is a process where the scatted photon gains energy from the system by de-exciting an initially free electron to a bound state.
% Let us next consider the bold solid red curve, that shows forward scattering data that have been collected during the same experiment, i.e., probing the same conditions.
% In this case, we \textcolor{red}{[sth about physisc/Penn gap, and maybe SIF?]...}
% \todo{Forward scattering stuff}
% Most importantly, we find the same general trend on the up-shifted part of the spectrum as for $\theta=160^\circ$.
The intensity measured on the forward detector is indicated by the blue line in the same plot, averaged over 7 shots.
We observe the same general trend on the up-shifted part of the spectrum in the collective regime as for $\theta=160^\circ$.
Heuristically, this indicates a consistent signature of the hot electrons upon probing the system on substantially different length scales.
The light blue line in Fig.~\ref{fig:all-spectra} shows a similar spectrum averaged over significantly more individual runs for the second dataset, nominally taken at the same conditions.
The SIF plotted here in lightgreen corresponds to the one measured for the second dataset and averaged over the corresponding shots.
For both the forward and backward analysis on the first dataset, a similar SIF was used taking into account the relevant shots.
Since there was no backscattering signal recorded in the second dataset, we instead focus on analysing the forward scattering in greater detail.
% As is shown in the averaged spectum displayed (light blue), the higher statistics in this dataset give access to a much higher dynamic range, thus allowing a more detailed analysis than previously possible.

\begin{figure}
    \centering
    \includegraphics[width=0.5\textwidth,keepaspectratio]{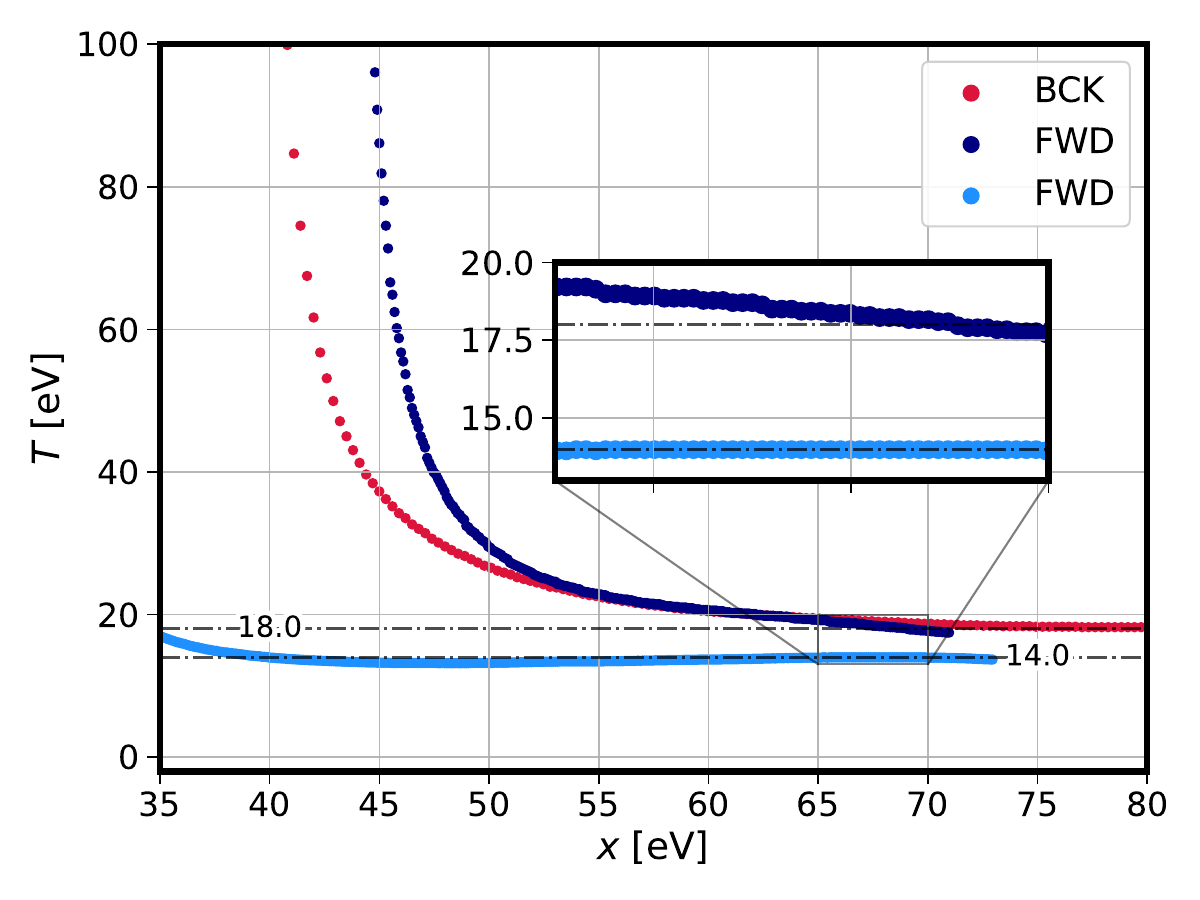}
    \caption{Inferred temperature as a function of the symmetrically truncated integration range $x$ for simultaneous backward and forward measurements for spectra shown in Fig.~\ref{fig:all-spectra}.
    The dark blue FWD line corresponds to the spectrum taken for dataset 1, whereas the light blue line is the forward data taken from the second dataset.
    }
    \label{fig:285-temperature-analysis}
\end{figure}

In Fig.~\ref{fig:285-temperature-analysis}, we show the model-free temperature extraction based on Eq.~(\ref{eq:symmetry}) for both forward and backward scattering intensities.
In particular, we plot the temperature derived using the minimum of the Laplace transform (i.e., $\tau=\beta/2=1/2T$) as a function of the symmetrically truncated integration range.
This is necessary as Eq.~(\ref{eq:Laplace}) assumes an infinite integration interval, whereas the available spectral range on a detector is always finite.
First consider the simultaneous forward and backward temperature analysis, shown in the blue and red curves in Fig.~\ref{fig:285-temperature-analysis}.
The effect of the SIF is taken into account here by convolving the measured source signal with an asymmetric HAPG response.
For more details on the construction of the SIF, see Ref.~\cite{gawne_SIF}.
While the inferred temperature appears to converge around $x=65\,$eV for the BCK signal (red), the forward signal (dark blue) shows a similar trend but does not converge as clearly.
This gives an estimate of the temperature for $T=18\,$eV for both, although the forward scattering appears significantly less converged and certain.
In comparison, looking at dataset 2 in Fig.~\ref{fig:285-temperature-analysis} (light blue), which contains higher statistics, a significantly clearer temperature convergence is observed at $x=50\,\unit{eV}$, giving a best estimate for the temperature at $T=14\,\unit{eV}$.

\begin{figure}
    \centering
    \includegraphics[width=0.49\textwidth]{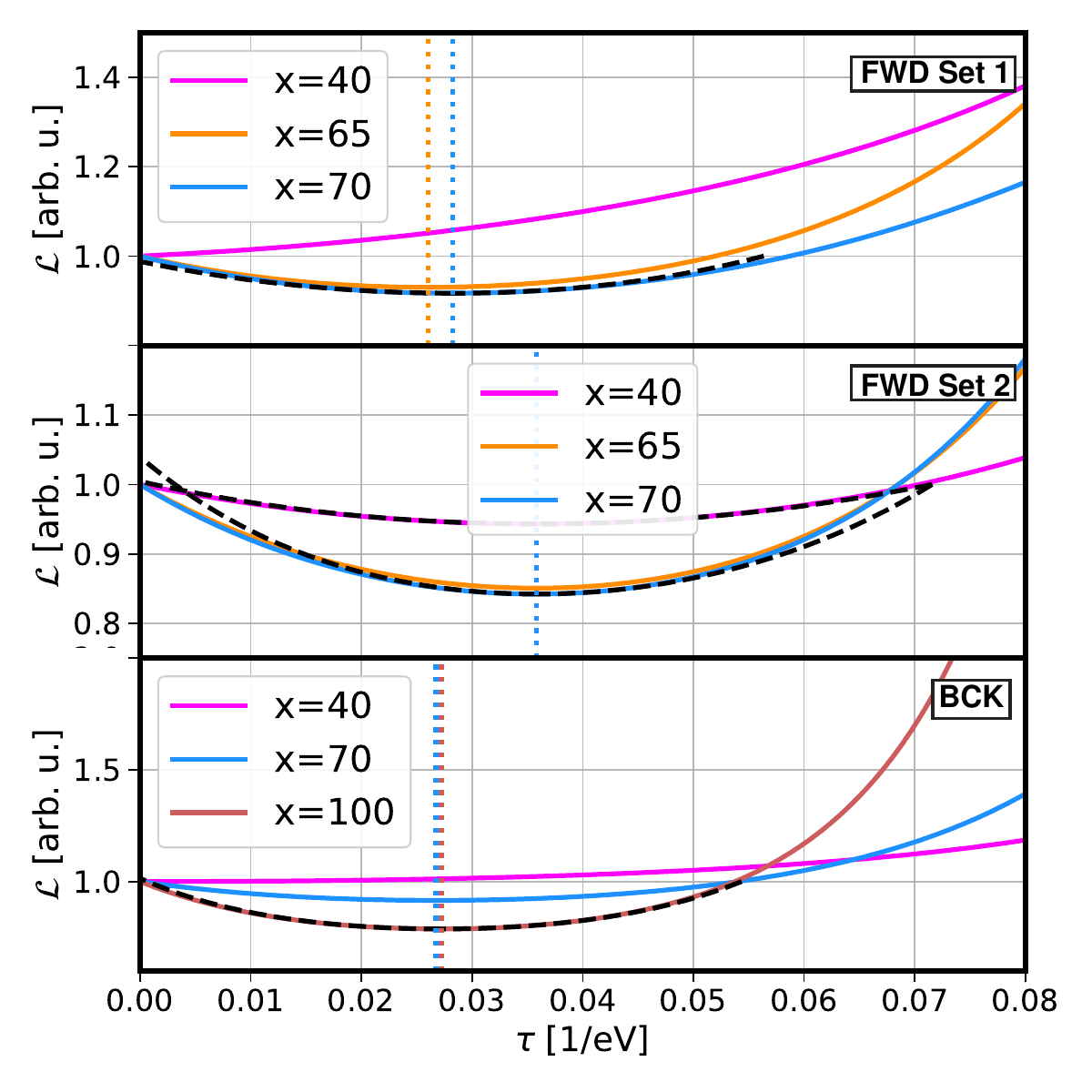}
    \caption{Deconvolved Laplace transform [cf.~Eq.~(\ref{eq:ITCF})] of the forward scattering intensity for two different symmetrically truncated integration intervals $x$. The corresponding dashed black lines have been mirrored around the minimum $\tau=\beta/2$ according to the relation $F(q,\tau)=F(q,\beta=\tau)$, cf.~Eq.~(\ref{eq:symmetry}).
    Minima are indicated using the dotted horizontal lines.
    }
    \label{fig:ITCF}
\end{figure}

Let us conclude our investigation by actually investigating the deconvolved Laplace transform [Eq.~(\ref{eq:ITCF})] of the forward scattering intensity in Fig.~\ref{fig:ITCF} for each dataset.
Corresponding results for the backscattering data have already been presented in Ref.~\cite{Dornheim_T2_2022} and are here repeated for completeness in the bottom panel.
All curves have been computed for different symmetrically truncated integration intervals, and have been normalized arbitrarily to $F(q,0)\equiv1$.
We note that it is in principle possible to extract the proper normalization of the ITCF from the f-sum rule in the imaginary-time domain~\cite{dornheim2023xray}, but this would require a spectral range of $I(q,E)$ beyond the K-edge.
First, we note for the forward analysis in the top panel, that the ITCF has evidently not converged for $x=60\,\unit{eV}$ or $x=70\,\unit{eV}$, and the position of its minimum is shifting to larger values (i.e. lower temperatures).
This confirms observations from Fig.~\ref{fig:285-temperature-analysis}.
In contrast, the ITCF plotted at different integration limits for the backward spectrum (bottom panel) has very clearly converged to a temperature, as the position of the minimum does not significantly shift for increasing $x$.
For the middle panel, representing the second dataset in the collective regime, we similarly see a converged temperature with increasing $x$.
Secondly, considering the mirrored ITCF indicated by the dashed black lines for each case, one can clearly see that Eq.~(\ref{eq:symmetry}) is noticeably violated for the middle panel, whereas it seems to hold in the non-collective regime represented in the bottom panel.
Notably, the asymmetry observed appears to increase with increasing $x$.
On second thought, $F(q,\tau)$ for $\tau>\beta/2$ is predominantly shaped by the up-shifted side of the scattering intensity (see Ref.~\cite{Dornheim_T2_2022} for an extensive discussion of the underlying mechanics), which exhibits a sharp drop around $\omega=60\,$eV.
We also note that any uncertainty for large $\tau$ is likely amplified by uncertainties in the SIF, making the symmetry of $F(q,\tau)$ investigated in Fig.~\ref{fig:ITCF} a less robust criterion for non-equilibrium compared to the comparison of temperatures extracted from different scattering angles as it is shown in Fig.~\ref{fig:285-temperature-analysis}.

\textbf{Discussion.} % \todo{Hannah: also needs to be re-written}
In this work, we have presented a model-free analysis of an XRTS measurement of isochorically heated warm dense graphite at two independent scattering angles.
From a physical perspective, probing the systems on the correspondingly different length scales offers a gamut of advantages.
First, XRTS measurements at different $q$ are likely sensitive to different parameters.
This fact can be exploited in future works to constrain the uncertainty in forward models~\cite{Kasim_POP_2019}.
Moreover, having independent data describing the same system allows one to check the consistency of the inferred parameters, and to learn more about a-priori hidden uncertainties. 
The latter point is particularly valuable to detect possible signatures of non-equilibrium, which is a serious concern for self-scattering experiments.
In the present case, we find an overall good consistency regarding the extracted temperatures, with $T\sim18\,$eV in both scattering directions for measurements taken simultaneously.
A second dataset at nominally the same conditions but analysed only for the forward scattering due to experimental constraint has given an estimated temperature of the same order of magnitude around $T\sim14\,\unit{eV}$.
The analysis has further shown that the dynamic range of the measured spectra and SIF significantly constrains the convergence of the ITCF method, particularly in the forward direction.
An accurate temperature in this regime could only be derived for a dataset averaged over a larger number of individual shots.
Potential signatures of non-equilibrium, which do in principle also manifest as an asymmetry in the converged ITCF,
are likely hidden either by the uncertainty in the SIF, or by the limited available dynamic range.%, but can be investigated by considering the symmetry of the converged ITCF.

The present proof-of-principle analysis has demonstrated the practical importance of probing multiple scattering angles for the rigorous characterization of extreme states of matter.
We are convinced that improved experimental set-ups that are based on existing capabilities will allow one to extract system parameters with even higher accuracy, and, in particular, to more reliably resolve even small signatures of non-equilibrium in the measured scattering intensity~\cite{Vorberger_PRX_2023}.
An important component of these efforts will likely be given by the utilization of seeded and monochromated x-ray beams (and potentially diced crystal analyzers~\cite{Gawne_PRB_2024}), which, in combination with improved SIF characterization, will reduce the uncertainties due to the source function.
This will be beneficial both for the detection of non-equilibrium, and for the resolution of lower temperatures in EOS measurements~\cite{Dornheim_T2_2022}.
In addition, we advocate for the measurement of a broad spectral range, which is important to extract additional observables such as the electronic static structure factor~\cite{dornheim2023xray} and the Rayleigh weight~\cite{dornheim2024modelfreerayleighweightxray}.
Finally, we stress the importance of a large dynamic range, which, usually, is the limiting factor when it comes to the model-free extraction of lower temperatures (i.e., $T\sim1\,$eV).
In addition to more reliable EOS measurements, such improved capabilities would also allow for the model-free study of relaxation rates at modern XFEL facilities with unprecedented resolution.

%We note that while we are looking at older experimental data here,
Indeed, many of the required experiment capabilities outlined above are now routinely available at XFEL facilities. For example, the large detection area and dynamic gain switching of Jungfrau detectors~\cite{Jungfrau_Detector} provides both a large dynamic range (with single photon sensitivity) in single shots, and a large spectral range when coupled with commonplace dispersive crystals, e.g. highly annealed/oriented pyrolytic graphite~\cite{Preston_JoI_2020}. Furthermore, the recent introduction of the DiPOLE-100X~\cite{DiPOLE100} and ReLaX~\cite{ReLaX} laser systems at the HED Scientific Instrument at the European XFEL~\cite{Zastrau_HED} now allows for HED systems to be reproducibly produced at repetition rates of 1-10~Hz -- a rate consistent with the probing FEL -- allowing for the rapid accumulation of data. This latter point is critical for the improvement of dynamic range as it allows for averaging of spectra from the same system over a large number of shots, thereby improving the sampling of spectral regions in the single photon regime.
We therefore anticipate that in the near future that there will be exciting developments in the model-free interpretation of XFEL experiments.

\section*{Acknowledgments}
This work was partially supported by the Center for Advanced Systems Understanding (CASUS) which is financed by Germany’s Federal Ministry of Education and Research (BMBF) and by the Saxon state government out of the State budget approved by the Saxon State Parliament. This work has received funding from the European Union's Just Transition Fund (JTF) within the project \emph{R\"ontgenlaser-Optimierung der Laserfusion} (ROLF), contract number 5086999001, co-financed by the Saxon state government out of the State budget approved by the Saxon State Parliament.
This work has received funding from the European Research Council (ERC) under the European Union’s Horizon 2022 research and innovation programme (Grant agreement No. 101076233, "PREXTREME"). 
Views and opinions expressed are however those of the authors only and do not necessarily reflect those of the European Union or the European Research Council Executive Agency. Neither the European Union nor the granting authority can be held responsible for them.
The work of H.M.B., B.B., M.P.B. Ti.D., M.J.M. was performed under the auspices of the U.S. Department of Energy by Lawrence Livermore National Laboratory under Contract No. DE-AC52-07NA27344 and supported by Laboratory Directed Research and Development (LDRD) Grants No. 24-ERD-044 and 25-ERD-047.
%The PIMC calculations were partly carried out at the Norddeutscher Verbund f\"ur Hoch- und H\"ochstleistungsrechnen (HLRN) under grant shp00026, and on a Bull Cluster at the Center for Information Services and High Performance Computing (ZIH) at Technische Universit\"at Dresden.

%%%%%%%%%%%%%%%%%%%%%%%%%%%%%%%%%%%%%%%%%%%%%%%%%%%%%%%%%%%%%%%%%%%%%%%%%%%%%%%%
% literature
%%%%%%%%%%%%%%%%%%%%%%%%%%%%%%%%%%%%%%%%%%%%%%%%%%%%%%%%%%%%%%%%%%%%%%%%%%%%%%%%
\bibliography{bibliography}
\clearpage
\end{document}